\documentclass{ifacconf}

\usepackage{graphicx}      % include this line if your document contains figures
\usepackage{natbib}        % required for bibliography
\usepackage{amsmath, amssymb, mathtools}
\usepackage{color}
\usepackage{subfig} 
\usepackage{comment}

\usepackage{url}

\newcommand{\COtwo}{\mathrm{CO}_2}

\newcommand{\degrees}{^\circ\mathrm{C}}
\newcommand{\Tindoor}{\mathrm{T}_\mathrm{indoor}}

\newcommand{\RH}{\mathrm{RH}}

\begin{document}
\begin{frontmatter}

\title{Generalizability of Learning-based Occupancy Detection in Residential Buildings (extended version)\thanksref{footnoteinfo}} 

\thanks[footnoteinfo]{This work was supported by the Swedish Energy Authority and IQ Samhällsbyggnad project DOCENT (project number P2023-01513, agreement 2023-205321), by the Digital Futures project HiSSx, and partially supported by the Wallenberg AI, Autonomous Systems and Software Program (WASP) funded by the Knut and Alice Wallenberg Foundation.}

\thanks[df]{Also affiliated with Digital Futures, Stockholm, Sweden.}
\author[Shared_1]{M. Farjadnia,} 
\author[Shared_2]{K. Eshkofti\thanksref{df},} 
\author[Shared_2]{A. Apell,} 
\author[Shared_2]{T. Hjalmarsson,} 
\author[Shared_2]{K. H. Johansson\thanksref{df},}
\author[Shared_2]{A. Fontan\thanksref{df},}
\author[Shared_1]{M. Molinari\thanksref{df}}
\address[Shared_1]{Department of Energy Technology,}
\address[Shared_2]{Department of Decision and Control Systems,\\ KTH Royal Institute of Technology, 100 44 Stockholm, Sweden,\\
E-mail:  \{mahsafa,eshkofti,kallej,angfon,marcomo\}@kth.se.}

\begin{abstract}                
This paper investigates non-intrusive occupancy detection methods for residential buildings using environmental sensor data from the KTH Live-In Lab in Stockholm, Sweden. Three machine learning approaches, namely, logistic regression (LR), support vector machines (SVM), and long short-term memory (LSTM) network enhanced with an attention mechanism, are evaluated in terms of predictive performance and computational complexity. The analysis considers the trade-off between sensor availability (investment cost) and prediction accuracy in real applications, as well as the models’ cross-apartment generalizability. Hyperparameters for both the SVM and LSTM models are optimized using Bayesian optimization. All three models are evaluated on data collected from apartments not used during training, and on data generated from a calibrated digital model of the testbed. Results show that all models achieve comparable performance on the same-apartment test data (accuracy $\approx$ 0.83, F1 score $\approx$ 0.86). When assessed on cross-apartment data, the LSTM model demonstrates the strongest generalization capability (accuracy of 0.84, F1 score of 0.85), while LR provides a competitive, low-complexity alternative for applications that do not require cross-apartment generalization.
\end{abstract}


\end{frontmatter}
%===============================================================================

\section{Introduction}
Buildings in the European Union contribute approximately 30$\%$ of the total final energy use and 26$\%$ of greenhouse gas emissions \citep{IEA_Building}. In this sector, heating, ventilation, and air conditioning (HVAC) systems are among the main drivers of energy use \citep{drgovna2020all}, and their performance strongly depends on occupant presence and behavior \citep{xu2023critical}. As occupant-centric control strategies continue to gain attention, studies have shown that accurate occupancy information can improve both energy efficiency and indoor comfort \citep{rueda2020comprehensive,park2019critical}. The growing demand for reliable occupancy estimation, coupled with increased digitalization of the building environment, has in turn driven extensive research into machine learning approaches for occupancy detection.
 
While occupancy detection has been widely studied (see review by \citep{jin2021building}), most studies focus on academic and office buildings, where occupants follow relatively regular schedules \citep{amiri2025review}. Residential buildings, however, have greater behavioral variability and face higher privacy constraints, resulting in comparatively limited research attention. Existing studies distinguish between intrusive sensing methods, such as cameras, and non-intrusive, low-cost, and privacy-preserving alternatives, including passive infrared (PIR), indoor temperature, humidity, and $\COtwo$, which are more suitable for residential buildings \citep{azimi2022fit,andersen2024exploring}.

A variety of machine learning techniques have been applied to occupancy detection \citep{rueda2020comprehensive}, ranging from traditional models such as logistic regression (LR) \citep{shi2017energy,chen2017environmental}, support vector machines (SVM) \citep{wang2018occupancy,khalil2021transfer}, and hidden Markov model (HMM) \citep{candanedo2017methodology}, to more advanced deep learning approaches, including feed-forward neural networks (FNNs) \citep{dong2010information} and long short-term memory (LSTM) networks \citep{kim2019real,khalil2021transfer}. Notably, \citep{chen2017environmental} developed occupancy prediction models for an office building using $\COtwo$, indoor temperature, and relative humidity (RH), with ground-truth occupant counts obtained from cameras. The proposed HMM with multinomial logistic regression outperformed the standard HMM. \citep{wang2018occupancy} compared SVM, K-nearest neighbors, and artificial neural network (ANN) for occupancy detection using $\COtwo$, indoor temperature, RH, and Wi-Fi signal data in a graduate student office, with camera-recorded ground truth. Their results showed that the ANN model achieved the highest accuracy when both environmental and Wi-Fi data were used, while SVM performed best when using Wi-Fi data alone. \citep{kim2019real} proposed an LSTM-based occupancy prediction model for individual zones of a large exhibition hall, using zone-level occupant counts obtained from image and optical sensors. The results demonstrated that LSTM can achieve superior predictive performance compared with other models, including the autoregressive integrated moving average (ARIMA). \citep{liang2024low} developed a non-intrusive occupancy estimation approach using $\COtwo$, indoor temperature, and RH collected from two university classrooms. The authors compared several machine learning models and found that LSTM provided the most accurate predictions, highlighting the feasibility of environmental sensor-based occupancy estimation without relying on intrusive ground-truth data. However, these studies rely on office and academic settings or on intrusive ground-truth methods such as cameras, and their generalizability to residential settings remains underexplored.

In residential settings, \citep{li2017new} studied short-term occupancy forecasting using data from four houses using a moving-window inhomogeneous Markov model and compared it with other methods, such as ANN, while \citep{huchuk2019comparison} compared multiple machine-learning models for future occupancy prediction using connected thermostat data from residential homes. However, explicit evaluation of transfer to unseen dwellings was not the primary focus of these studies.

This paper addresses this gap by focusing on occupancy detection in residential buildings using non-intrusive indoor environmental data collected from the KTH Live-In Lab in Stockholm, Sweden. We apply three machine learning approaches, two widely used methods (LSTM and SVM) and a lower-complexity alternative (LR), to evaluate their predictive performance and computational complexity. We also compare their performance under varying sensor-availability (investment cost) conditions to examine how well the models operate when only a limited set of sensor types is available in practice. Finally, we assess cross-apartment generalizability by testing the models on data from previously unseen apartments in the KTH Live-In Lab, as well as on data generated from a calibrated digital model of the testbed.

This paper is structured as follows. Section~\ref{sec:Expermental-setup} introduces the experimental setup and the dataset collected from the KTH Live-In Lab. Section~\ref{sec:Method} outlines the learning techniques used to derive the occupancy detection models based on the dataset.  Section~\ref{sec:Results} presents the results, and finally, Section~\ref{sec:Conclusion} provides concluding remarks.

\section{Experimental Setup and collected dataset}\label{sec:Expermental-setup}

This section introduces the building environments utilized in this study: the KTH Live-In Lab and Testbed KTH (Section~\ref{sec:KTH-lil}), as well as the dataset collected for occupancy detection (Section~\ref{sec:KTH-dataset}). 

\subsection{The KTH Live-In Lab}\label{sec:KTH-lil}

\begin{figure*}[th!]\centering
\subfloat[]{\includegraphics[height=4.1cm,trim={0 0 0 0.5cm},clip]{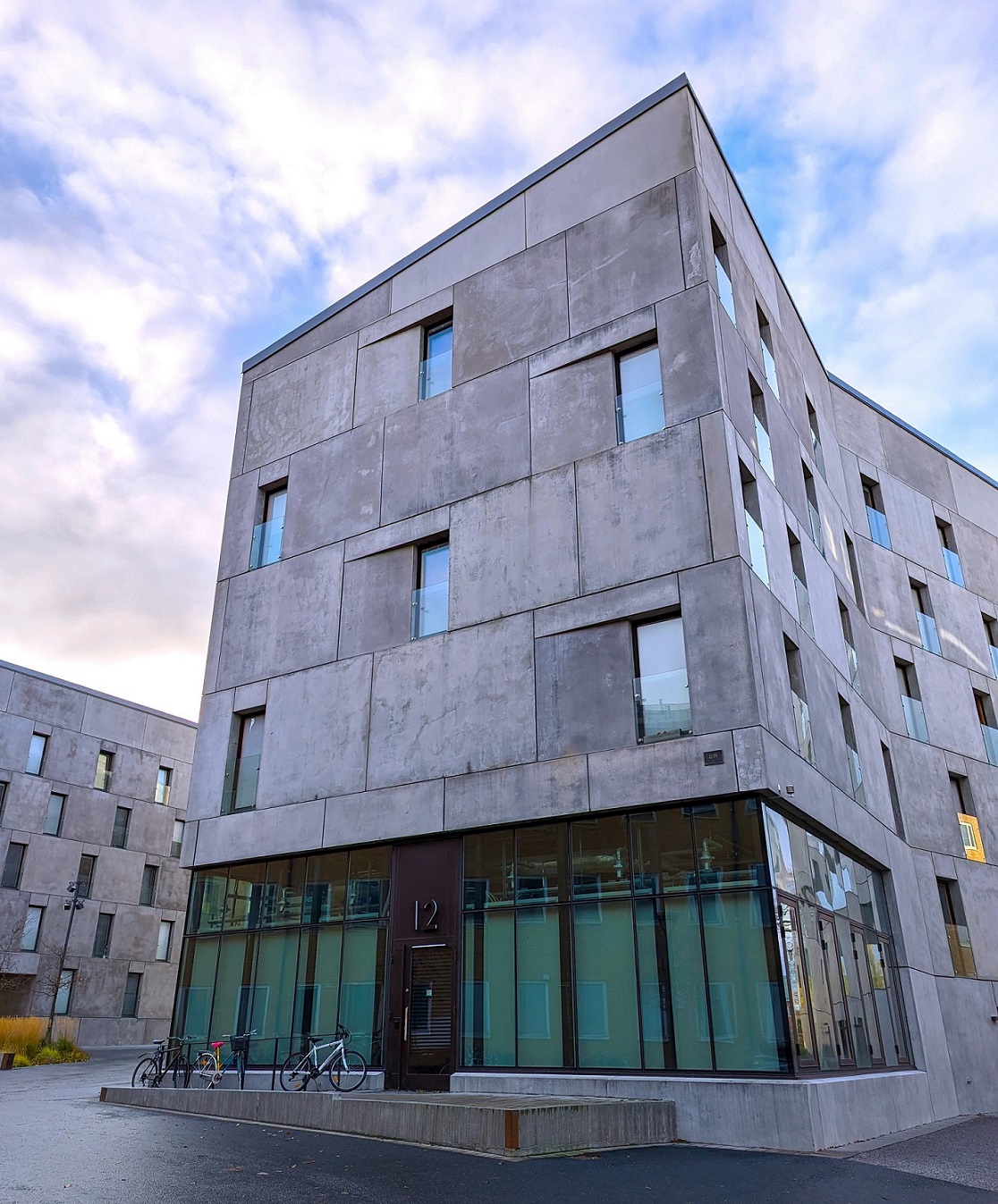}\label{fig:Outside}}
% ...................................
\quad \quad
% ...................................
% ...................................
\subfloat[]{\includegraphics[height=4cm]{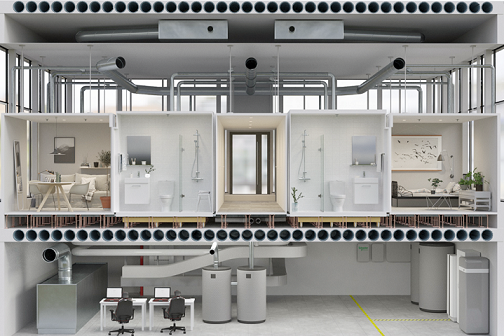}\label{fig:sectional-view}}
\quad \quad
\subfloat[]{\includegraphics[height=4cm]{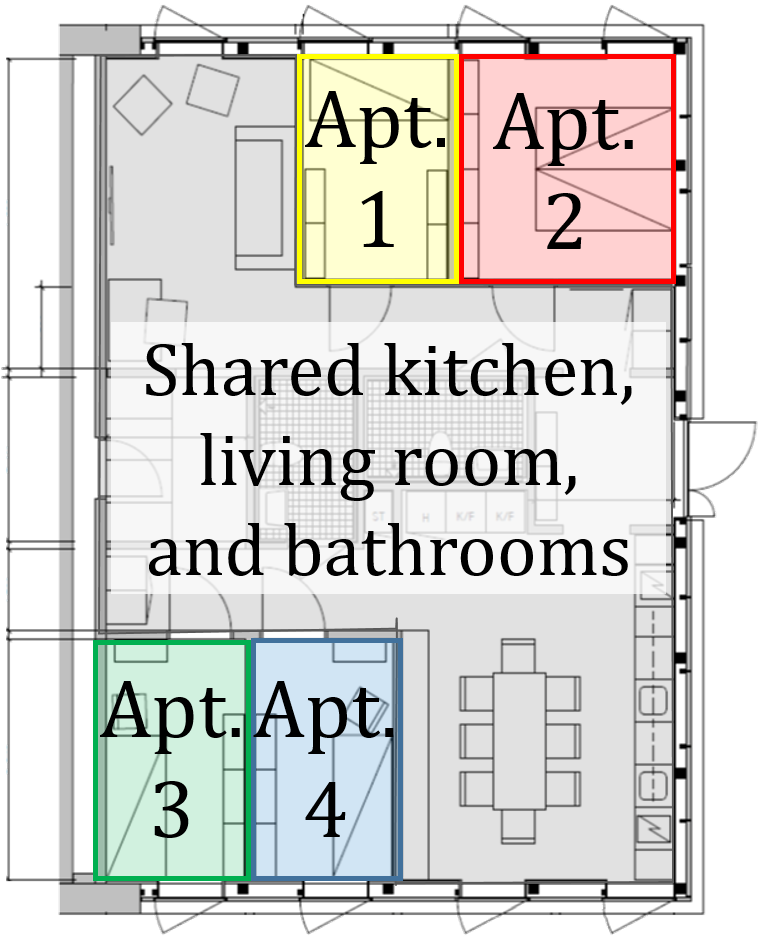}\label{fig:Testbed3}}
% ...................................
\quad \quad
\caption{(a): Exterior view of Testbed KTH. (b): Sectional illustration of Testbed KTH.  (c): Latest layout of Testbed KTH.}
%\url{https://www.liveinlab.kth.se/en/infrastruktur/testbed-infrastructure}
\label{fig:TestbedEM}
\end{figure*}

The KTH Live-In Lab\footnote{\url{https://www.liveinlab.kth.se/en}.} comprises a variety of building testbeds, including student housing and lecture halls \citep{molinari2023using}. This study focuses on one of these facilities, known as Testbed KTH. Testbed KTH is equipped with an extensive sensing infrastructure that continuously records indoor environmental parameters, such as indoor temperature, relative humidity, $\COtwo$ concentration, and volatile organic compounds, as well as energy use for space heating and domestic hot water production. In addition, magnetic sensors track the open or closed status of windows and doors, while PIR presence detectors monitor occupant presence. PIR sensors are typically considered privacy-preserving (non-intrusive) sensors in residential settings, particularly when compared with camera-based sensors.  
Further details can be found in \citep{molinari2023using,rolando2022long}. Testbed KTH covers a total floor area of 300$\mathrm{m}^2$ (Figs.~\ref{fig:Outside} and ~\ref{fig:sectional-view}), and it accommodates undergraduate students (of comparable age) from the KTH Royal Institute of Technology who live there full-time. Each apartment is typically rented to a single occupant, with one apartment being rented out to one to two occupants.

\subsection{Experimental dataset}\label{sec:KTH-dataset}
The dataset utilized in this work includes measurements collected between June 2022 and November 2024 from Apartment 2 in Testbed KTH. During this period, the apartment was occupied by different tenants, each holding approximately yearly lease contracts. The sensors installed in the apartment are standard commercial devices commonly used in building monitoring applications. Their measurement accuracy is consistent with standard performance levels reported in the literature \citep{cali2016analysis}; for instance, the indoor and outside temperature sensors have accuracies of $\pm 0.5\degrees$ and $\pm 0.6\degrees$, respectively. Indoor temperature, relative humidity, and $\COtwo$ concentration were measured using a wall-mounted sensor installed at approximately 1.6m and positioned away from windows, doors, direct sunlight, ventilation outlets, and exterior walls. Given the small apartment size, the $\COtwo$ concentration was assumed to be reasonably well mixed within the monitored space. Therefore, the measurements were considered representative of indoor conditions.

\begin{figure*}[t!]\centering
\subfloat{\includegraphics[width=0.33\linewidth]{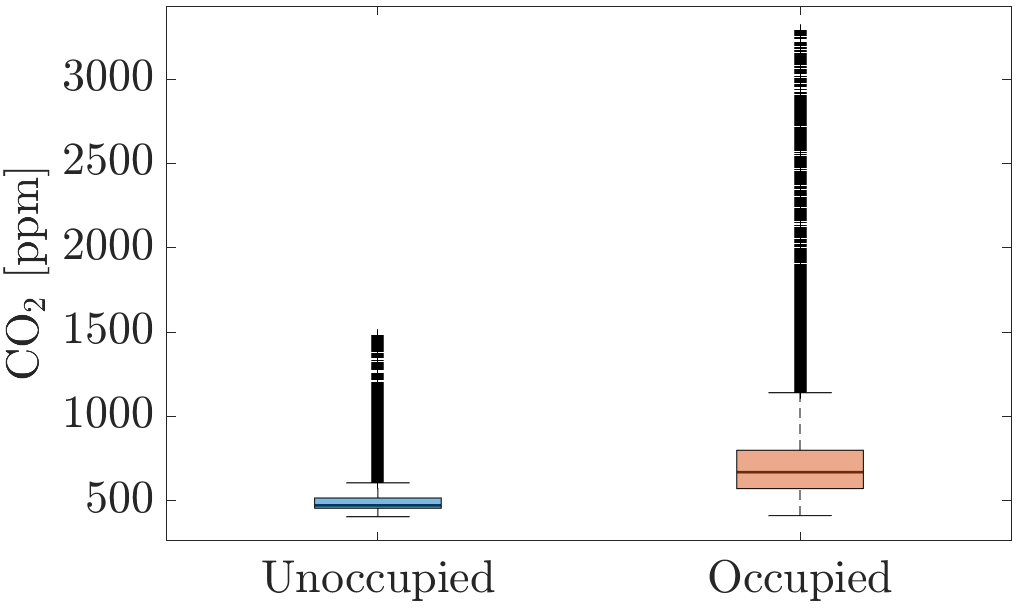}\label{fig:CO2_box}}
% ...................................
\subfloat{\includegraphics[width=0.33\linewidth]{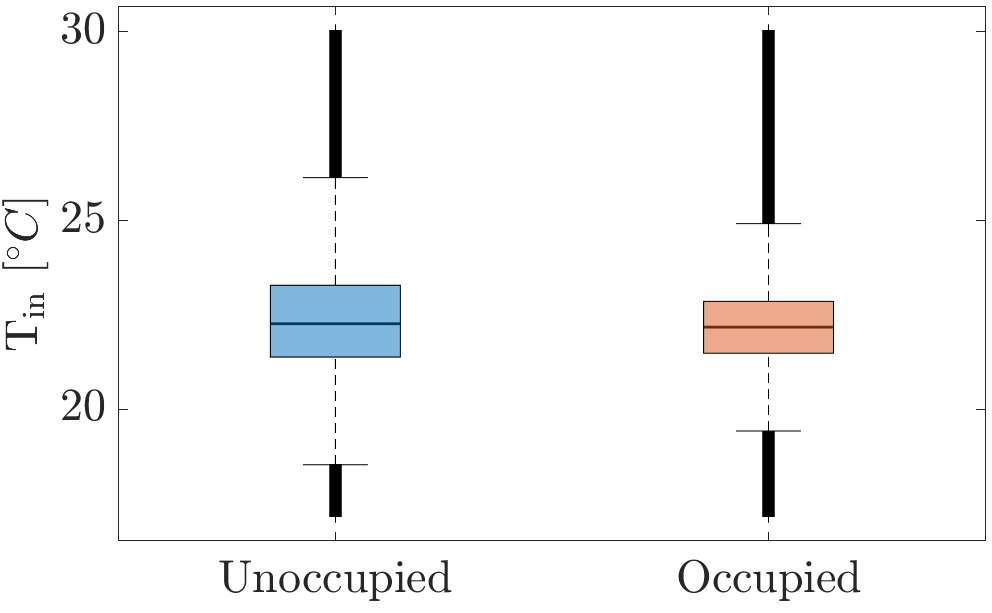}\label{fig:Tin_box}}
\subfloat{\includegraphics[width=0.33\linewidth]{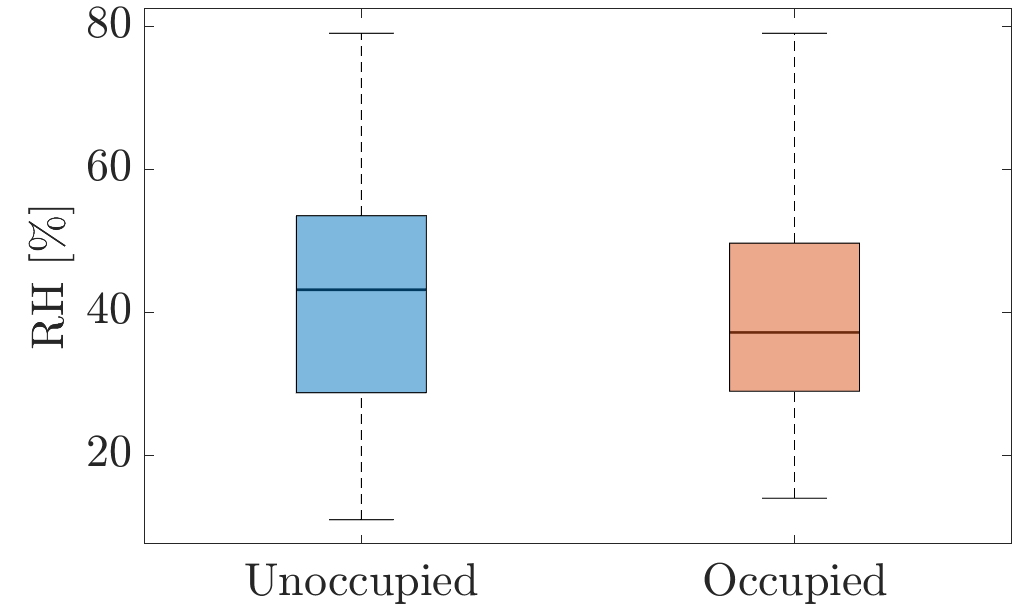}\label{fig:RH_box}}
% ...................................
\caption{Boxplots of monitored variables in Apartment 2 during occupied and unoccupied periods.}
\label{fig:Data_boxplots}
\end{figure*}

\begin{figure}
    \subfloat{\includegraphics[width=\linewidth]{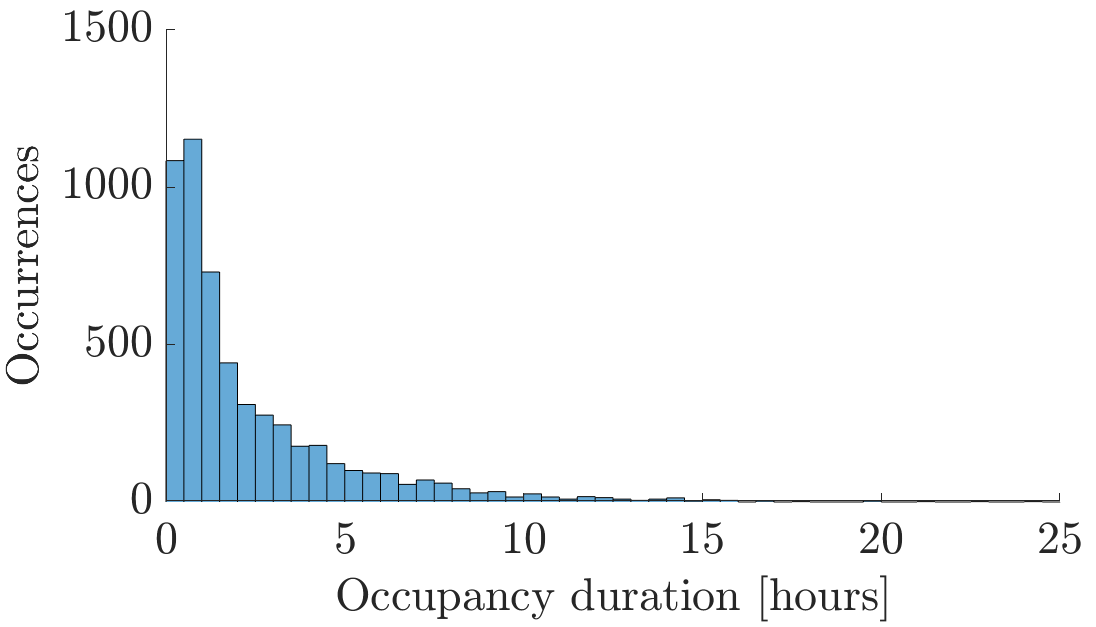}}
    \caption{Distribution of occupancy durations recorded in Apartment 2 from June 2022 to November 2024.}
    \label{fig:occ_his}
\end{figure}

\begin{figure}\centering
\subfloat[]{\includegraphics[width=\linewidth]{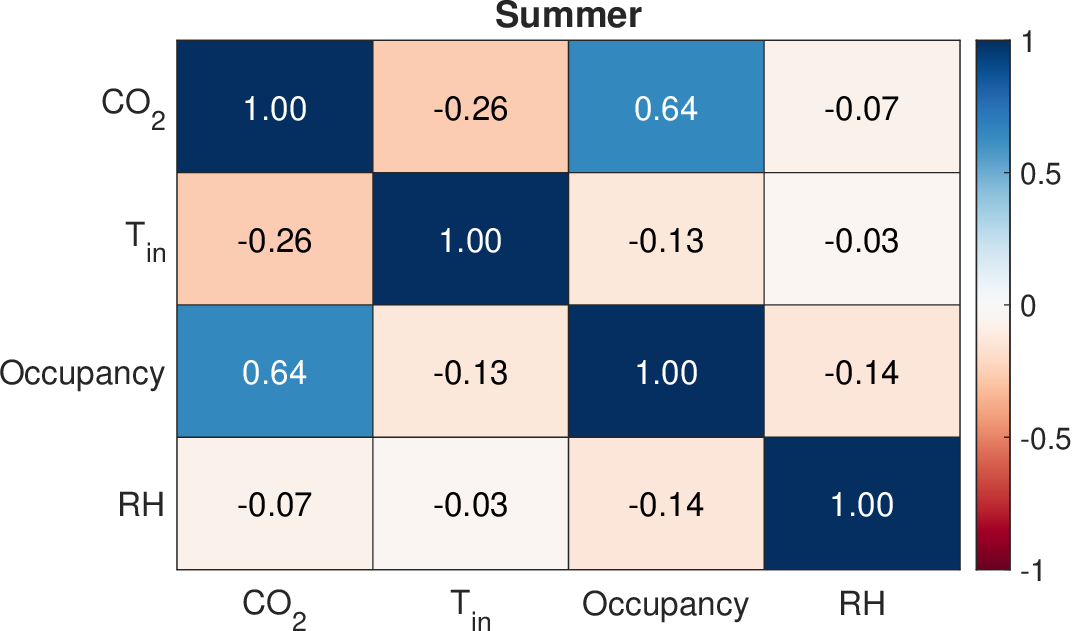}\label{fig:corr_summer}}\\
% ...................................
\subfloat[]{\includegraphics[width=\linewidth]{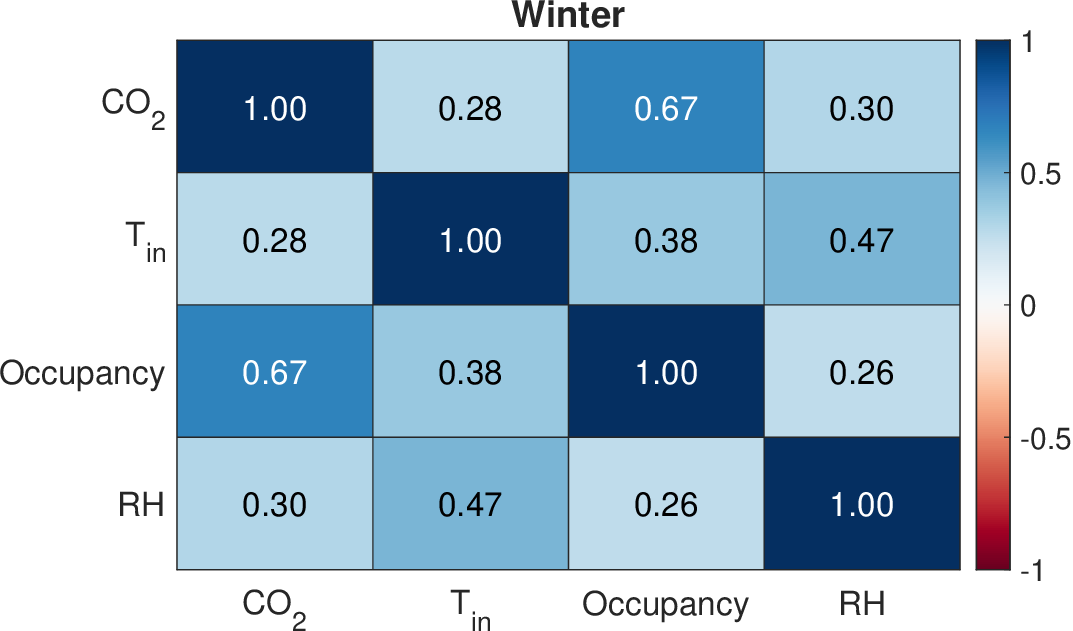}\label{fig:corr_winter}}
% ...................................
\caption{Correlation matrices of environmental variables and occupancy in Apartment 2 for (a) the summer period and (b) the winter period, covering June 2022 to November 2024.}

\label{fig:correlation}
\end{figure}

The data were preprocessed before use to handle missing values, remove incorrect measurements, and account for discrepancies arising from nonuniform sampling rates across sensors. Following these procedures, the variables listed below were considered with a sample time of 30 seconds, see also Figs.~\ref{fig:Data_boxplots} and~\ref{fig:occ_his} for illustration:
%\begin{comment}
\begin{itemize}
\item Occupancy status (vacant/occupied);
\item Indoor temperature ($\Tindoor$) $[\degrees]$;
\item Carbon dioxide concentration ($\COtwo$) [ppm];
\item Relative humidity (RH) $[\%]$.
\end{itemize}
The variation in $\COtwo$ between occupied and unoccupied periods is considerably larger than that observed for $\Tindoor$ and $\RH$ (Fig.~\ref{fig:Data_boxplots}). This dynamic is expected to be more noticeable in small, single-occupant apartments, where $\COtwo$  is more directly influenced by human presence and fresh-air exchange, whereas temperature and humidity tend to vary more slowly due to thermal mass and HVAC regulation.
The distribution of occupancy durations is presented in Fig.~\ref{fig:occ_his}. Most occupied intervals last less than one hour, indicating that the occupant frequently leaves the apartment for short periods. Because the correlations between occupancy and environmental variables are season-dependent, driven by changes in HVAC control strategies, window-opening behavior, and outdoor conditions, we analyzed the correlation matrices separately for different seasons. This approach allows us to capture seasonal differences in feature correlations that may influence the performance of the models. Two representative seasonal correlation matrices are shown in Fig.~\ref{fig:correlation}.

\section{Methodology}\label{sec:Method}
This section outlines the methodological framework used to develop and evaluate learning-based occupancy detection models. First, the occupancy prediction problem is formalized, followed by a description of the feature engineering process. Subsequently, three learning methods, namely LR, SVM, and LSTM, are described in detail, including their modeling assumptions, decision-rule formulations, and hyperparameter optimization procedures.
\subsection{Problem formulation}\label{sec:problem}
As detailed in Subsection~\ref{sec:KTH-dataset}, sensor measurements are sampled every $\Delta t = 30\,\mathrm{s}$. Let $t \in \{1, \dots, T\}$ index these discrete time steps. At each time step $t$, the binary occupancy state of the apartment is represented by $y_t \in \{0,1\}$, where $y_t = 1$ corresponds to an \emph{occupied} apartment and $y_t = 0$ to a \emph{vacant} one. The goal is to learn data-driven predictors that map sensor measurements to occupancy labels. Any such predictor can be expressed abstractly as
\[
\hat{y}_t = \mathcal{F}(u_t, \Theta),
\]
where $u_t$ denotes the model input at time $t$, and $\Theta$ represents the model-specific parameters. The specific form of $u_t$ depends on whether a static model, such as LR or SVM, or a sequential model, such as an LSTM, is employed, as discussed below. However, the label $y_t$ is identical for all three models. In LR and SVM, the input is a feature vector $\boldsymbol{x}_t \in \mathbb{R}^d$, where $d$ denotes the number of features. These models therefore learn a static decision function
\[
\hat{y}_t = \mathcal{F}_{\mathrm{static}}(\boldsymbol{x}_t, \Theta),
\]
where $\mathcal{F}_{\mathrm{static}}: \mathbb{R}^d \to [0,1]$ approximates either the posterior probability, as in LR, or a margin-based decision score, as in SVM, which will be explained in Subsections~\ref{sec:LR} and ~\ref{sec:SVM}, respectively.
On the other hand, LSTM models operate on histories of feature vectors to learn temporal dependencies. Therefore, given a sequence length $L_{\mathrm{seq}}$, the input at time $t$ is the matrix
\[
\boldsymbol{X}_t = \big[\boldsymbol{x}_{t-L_{\mathrm{seq}}+1}, \dots, \boldsymbol{x}_t \big] \in \mathbb{R}^{d \times L_{\mathrm{seq}}}.
\]
This representation shows that the stacked feature vectors form a sliding time window ending at time $t$. Subsequently, the mapping and occupancy probability estimation are performed, which will be detailed in Subsection~\ref{sec:LSTM}.

\subsection{Feature engineering}
The dataset is split into training, validation, and test sets using a strictly time-ordered strategy, where the earliest portion is used to train the models, the subsequent portion is used for validation and threshold selection, and the most recent portion is held out for the final test.

A moderate class imbalance between occupied and vacant states is observed. This imbalance can be addressed within the learning algorithms by employing class-dependent loss weights or misclassification penalties, as described in the corresponding model subsections. Additionally, all subsequent models share the same label definition and feature construction presented in this section.

Since the reaction of the CO$_2$ concentration to occupancy changes is more pronounced than that of temperature or relative humidity, the raw CO$_2$ level at time $t$ is complemented with an approximation of its short-term temporal slope. A window length of 15 minutes is used to compute the CO$_2$ slope, expressed in units of $\mathrm{ppm}/\mathrm{s}$.

Moreover, as shown in Fig.~\ref{fig:correlation}, environmental measurements and occupancy patterns depend on the season, since outdoor conditions, HVAC control strategies, and typical occupant behavior vary throughout the year. To provide the models with this contextual information, the month corresponding to each sample's timestamp is extracted and mapped to one of four meteorological seasons: winter (December--February), spring (March--May), summer (June--August), and autumn (September--November). Each season is then encoded as a one-hot vector. Collecting the above-mentioned measurements, the feature vector used for all models is
\begin{equation}
    \boldsymbol{x}_t =
    \begin{bmatrix}
        \mathrm{CO2}_t &
        T_t &
        \mathrm{RH}_t &
        \Delta \mathrm{CO2}_t &
        \mathrm{Season}_t
    \end{bmatrix}^\top.
\label{eq:feature}
\end{equation}

All continuous features are standardized prior to training. Let $\mu_j$ and $\sigma_j$ denote the empirical mean and standard deviation of the $j$-th feature computed only on the training set. Each feature is then transformed as $\tilde{x}_t^{(j)} = \frac{x_t^{(j)} - \mu_j}{\sigma_j}$, resulting in the normalized vector $\tilde{\boldsymbol{x}}_t \in \mathbb{R}^d$. The same affine transformation is applied to the validation and test data. For the LSTM model, normalization is performed on each per-feature time series before constructing the input sequences $\mathbf{X}_t$, leading to 
$\tilde{\boldsymbol{X}}_t = \left[ \tilde{\boldsymbol{x}}_{t-L_{\text{seq}}+1}, \ldots, \tilde{\boldsymbol{x}}_t \right]$.

This shared preprocessing pipeline ensures that all learning algorithms operate on a consistent representation of the sensor measurements while enabling a fair comparison between static and sequence-based methods. In addition to implementation on the reference apartment, the generalizability of the learned predictors to other data sources and apartments is investigated. Therefore, in the later sections, an evaluation protocol is introduced that considers both reduced sensor-availability scenarios and cross-apartment testing.
\subsection{LR model}\label{sec:LR}
In the LR approach, the probability of the apartment being occupied ($p_t$) at time step $t$ is modeled using the logit function as follows:
\begin{align}\label{eqn:probability}
    \ln \left( {\frac{p_t}{{1 - p_t}}} \right) = \alpha  + \sum_{j=1}^d \beta_j x_t^{(j)},
\end{align}
where $\alpha$ is the intercept term corresponding to the different seasons, and $\beta_j$ denotes the coefficient associated with the $j$-th feature $x_t^{(j)}$ \citep[Chapter~2.2]{hosmer2013applied}.

\subsection{SVM model}\label{sec:SVM}
To obtain a strong nonlinear foundation for binary occupancy detection, an SVM with a radial basis function (RBF) kernel is employed. The SVM maps the sensor measurements to the binary occupancy state using the feature vector defined in eq.~\ref{eq:feature}. In the SVM formulation, the occupancy labels are mapped to $\tilde{y}_t \in \{-1, +1\}$ by $\tilde{y}_t = 2 y_t - 1$. Considering the training set $\{(\tilde{\mathbf{x}}_t, \tilde{y}_t)\}_{t=1}^{T}$, the soft-margin SVM solves
\begin{equation}
\begin{aligned}
    \min_{\mathbf{w},\, b,\, \boldsymbol{\xi}} \quad 
        & \frac{1}{2}\|\mathbf{w}\|_2^2 
        + C_{+} \sum_{t:\,\tilde{y}_t = +1} \xi_t 
        + C_{-} \sum_{t:\,\tilde{y}_t = -1} \xi_t , \\
    \text{s.t.} \quad 
        & \tilde{y}_t \left( \mathbf{w}^\top \phi(\tilde{\mathbf{x}}_t) + b \right) 
            \ge 1 - \xi_t, 
            \qquad t = 1,\ldots,T, \\
        & \xi_t \ge 0, \qquad t = 1,\ldots,T,
\label{eq:SVM}
\end{aligned}    
\end{equation}
where $\phi(\cdot)$ denotes the implicit feature mapping induced by the kernel, and $\mathbf{w}$ and $b$ are the weight vector of the SVM and the bias term, respectively. The variables $\xi_t$ are slack variables. The coefficients $C_{+}$ and $C_{-}$ are class-dependent regularization parameters that assign larger misclassification penalties to the minority class, helping mitigate class imbalance between occupied and unoccupied samples. The RBF kernel is typically defined as
\[
K({\mathbf{\tilde{x}}}_t,{\mathbf{\tilde{x}}}_{t'}) = \exp\!\left( -\gamma \lVert {\mathbf{\tilde{x}}}_t - {\mathbf{\tilde{x}}}_{t'} \rVert_2^2 \right),
\qquad t, t' = 1, \ldots, T.    
\]
where $\gamma > 0$ controls the kernel bandwidth and the complexity of the decision boundary. The SVM defines a decision function in its dual form as
\[
    f_{\mathrm{svm}}(\mathbf{x}) 
    = \sum_{t=1}^{T} \alpha_t \tilde{y}_t \, K(\tilde{\mathbf{x}}_t, {\mathbf{\tilde{x}}}_{t'}) + b,
\]
where $\alpha_t \ge 0$ are the learned dual coefficients. Instead of using the default hard decision rule $\hat{y} = \mathbb{I}\{f_{\mathrm{svm}}(\mathbf{x}) \ge 0\}$,  
the signed margin $f_{\mathrm{svm}}(\mathbf{x})$ is treated as a continuous score, and a scalar decision threshold $\tau$ is optimized on the validation set. Accordingly, the decision rule is defined as
\[
    \hat{y}(\mathbf{x};\tau) =
    \begin{cases}
        1, & \text{if } f_{\mathrm{svm}}(\mathbf{x}) \ge \tau, \\
        0, & \text{otherwise}.
    \end{cases}
\]
To select the optimal decision threshold, $\tau$, the algorithm searches over a range defined by empirical percentiles of the validation scores and chooses the value that maximizes the F1 score of the positive class. It is worth mentioning that, to increase computational efficiency on long time series, the training set is optionally subsampled to a fixed maximum size using stratified sampling to preserve the class distribution.

The two main SVM hyperparameters $(C, \gamma)$ are tuned via Bayesian optimization, as explained in Section~\ref{sec:BO}. Once the optimal hyperparameters $(C^{\star}, \gamma^{\star})$ and the optimal threshold $\tau^{\star}$ are determined, a final SVM model is retrained on the combined training and validation sets. Its performance metrics are then reported on the test data using the fixed threshold~$\tau^{\star}$.

\subsection{LSTM model}\label{sec:LSTM}
Fixed-length input sequences of the normalized features are constructed using a sliding window of length $L_{\mathrm{seq}}$. The sequence $\mathbf{X}_t$ is then processed by a stacked LSTM with $n_L$ layers and hidden dimension $H$. The network therefore generates a sequence of hidden states defined as
\begin{equation}
    (\boldsymbol{h}_{t,1}, \ldots, \boldsymbol{h}_{t,L_{\mathrm{seq}}})
    = f_{\mathrm{lstm}}(\mathbf{X}_t; \theta),
    \qquad
    \boldsymbol{h}_{t,k} \in \mathbb{R}^{H},
    \label{eq:hidden}
\end{equation}
where $\theta$ denotes the LSTM parameters and $k$ indexes the time steps within the input window. Instead of relying solely on the final hidden state, a global attention mechanism is employed to learn a data-driven aggregation of the hidden states over time and to improve generalization. The attention scores are computed as
\[
    e_{t,k}
    = \mathbf{v}^{\top} \tanh\!\left( W \boldsymbol{h}_{t,k} \right),
\]
where $W \in \mathbb{R}^{H \times H}$ and $\mathbf{v} \in \mathbb{R}^{H}$ are trainable parameters. The attention weights are computed by normalizing the scores along the temporal dimension as
\begin{equation}
        \beta_{t,k}
    =
    \frac{\exp(e_{t,k})}
         {\sum_{j=1}^{L_{\mathrm{seq}}} \exp(e_{t,j})},
    \qquad k = 1, \ldots, L_{\mathrm{seq}},
    \label{eq:score}
\end{equation}
which satisfy $\beta_{t,k} \ge 0$ and $\sum_{k=1}^{L_{\mathrm{seq}}} \beta_{t,k} = 1$. The weights $\beta_{t,k}$ can be interpreted as the significance of the $k$-th time step in the input window for predicting the occupancy at time $t$.
Finally, a context vector $c_t \in \mathbb{R}^{H}$ is then constructed as the attention-weighted sum
\[
    \boldsymbol{c}_t
    = \sum_{k=1}^{L_{\mathrm{seq}}} \beta_{t,k} \, \boldsymbol{h}_{t,k}.
\]
To ensure stabilized training, $\boldsymbol{c}_t$ is subsequently passed through a layer-normalization operator \citep{ba2016layernormalization}, yielding $\tilde{\boldsymbol{c}}_t = \mathrm{LN}(\boldsymbol{c}_t)$, which
normalizes each feature to zero mean and unit variance followed by a
learnable affine transformation.
A two layer fully connected prediction head with rectified linear unit activations is applied on top of $\tilde{\boldsymbol{c}}_t$ to produce a scalar logit $s_t$ via
\[
\begin{aligned}
    \boldsymbol{z}_{1,t} &= \Phi(W_1 \tilde{\boldsymbol{c}}_t + \boldsymbol{b}_1), \\
    \boldsymbol{z}_{2,t} &= \Phi(W_2 \boldsymbol{z}_{1,t} + \boldsymbol{b}_2), \\
    s_t     &= \boldsymbol{w}_{\mathrm{out}}^{\top} \boldsymbol{z}_{2,t} + b_{\mathrm{out}},
\end{aligned}
\]
where $W_1 \in \mathbb{R}^{32 \times H}$,    $W_2 \in \mathbb{R}^{16 \times 32}$,    $w_{\mathrm{out}} \in \mathbb{R}^{16}$,
and the biases $b_1$, $b_2$, $b_{\mathrm{out}}$ are trainable parameters. The layer widths ($32$ and $16$ units) are architectural
hyperparameters selected based on validation performance to provide
sufficient prediction capacity.

A dropout rate $p_{\mathrm{drop}} \in (0,1)$ is applied to the LSTM outputs and between the dense layers. It randomly sets each activation to zero with probability $p_{\mathrm{drop}}$ and rescales the remaining activations. This serves as a stochastic regularizer that helps mitigate overfitting. The logit $s_t$ is converted to an occupancy probability using the logistic sigmoid function.

Similar to the SVM model, the loss incorporates a class-dependent regularization coefficient to address class imbalance.
Training is performed using mini-batch stochastic gradient descent with the Adam optimizer, a learning-rate scheduler, and an early-stopping criterion based on the validation area under the ROC curve (AUC).
The decision rule and the scalar decision threshold follow the same F1-based procedure described for the SVM model in Section~\ref{sec:SVM}.

\subsection{Training and evaluation procedure}\label{sec:training}
All models are first trained and evaluated on data from the same reference apartment, which is Apartment 2 shown in Fig.~\ref{fig:Testbed3}. The time series is split chronologically into three disjoint sets to prevent data leakage across time. That is, the training set contains all samples from \(1\) June \(2022\) \(00{:}00{:}00\) to \(31\) December \(2023\) \(23{:}59{:}30\), which results in \(1\,667\,520\) time steps. The validation set covers the subsequent period from \(1\) January \(2024\) \(00{:}00{:}00\) to \(30\) June \(2024\) \(23{:}59{:}30\), with \(524\,160\) records. The final test set consists of all remaining measurements from \(1\) July \(2024\) \(00{:}00{:}00\) to \(6\) November \(2024\) \(03{:}11{:}00\), yielding \(369\,023\) samples. 

As noted in earlier subsections, class imbalance is addressed in both SVM and LSTM through class dependent misclassification penalties and weighted binary cross entropy loss, respectively. For performance assessment, we report accuracy, precision, recall, F1 score, ROC AUC, and the confusion matrix on the test set. In occupancy detection, both missed occupied periods (false negatives) and unnecessary activation of building services (false positives) are undesirable. Therefore, the F1 score, which is the harmonic mean of precision and recall, is the primary evaluation metric because it penalizes models that perform well on only one of these aspects.

In addition, the hyperparameters for the SVM and LSTM models are optimized by maximizing the validation F1 score through Bayesian optimization, as detailed in Subsection~\ref{sec:BO}.
\subsection{Hyperparameter optimization}\label{sec:BO}
Bayesian optimization (BO) is employed to automatically tune the hyperparameters of the SVM and LSTM models by maximizing the validation F1 score of the positive class, treated as a black-box objective
\[
\boldsymbol{\lambda}^\star
= \arg\max_{\boldsymbol{\lambda} \in \Lambda} J(\boldsymbol{\lambda}),
\qquad
J(\boldsymbol{\lambda}) = F1^{\mathrm{val}}(\boldsymbol{\lambda}),
\]

Here, $\boldsymbol{\lambda}$ denotes the vector of hyperparameters and $\Lambda$ is the set of all allowed hyperparameter vectors. Additionally, $F1^{\mathrm{val}}(\boldsymbol{\lambda})$ is the F$1$ score on
the validation set obtained with hyperparameters. Following the standard BO procedure introduced by \citep{Snoek2012}, a probabilistic surrogate model is iteratively fitted to the observed pairs $\{(\boldsymbol{\lambda}_j, J(\boldsymbol{\lambda}_j))\}_{j=1}^i$,
where $i$ is the iteration index. 
In particular, an acquisition function
$a_i(\boldsymbol{\lambda})$, such as expected improvement, is then
maximized to select the next candidate
\[
\boldsymbol{\lambda}_{i+1}
= \arg\max_{\boldsymbol{\lambda} \in \Lambda}
a_i(\boldsymbol{\lambda}).
\]
Each proposed configuration $\boldsymbol{\lambda}_{i+1}$ is trained on
the training set and evaluated on the validation set. After a predefined number of evaluations, the best-performing hyperparameters are used to retrain the final model on the combined training and validation data. This mechanism allows BO to use the validation set specifically for model selection, while the final model is trained on all available labeled data except the test set, which is held out to provide an unbiased estimate of generalization performance.

\section{Results and Discussion}\label{sec:Results}

This section presents the evaluation of the LR, SVM, and LSTM models developed following the methodology outlined in Section~\ref{sec:Method}. In particular, Section~\ref{sec:performance} assesses the performance of all models using test data from the reference apartment. Section~\ref{sec:generalization} examines the models’ generalization capabilities by comparing their performance on synthetic data generated from a digital model of Testbed KTH and on the experimental data gathered from another apartment within Testbed KTH.

\subsection{Performance on reference apartment}\label{sec:performance}

\begin{figure}
    \centering
    \includegraphics[width=0.85\linewidth]{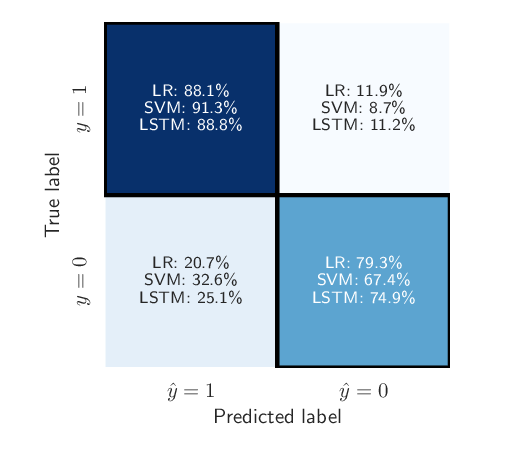}
    \caption{Row-normalized confusion matrices for the models evaluated using the Apartment 2 test dataset, with each cell showing the proportion of samples within the corresponding true-class row.}
    \label{fig:confusion_matrix}
\end{figure}

In this section, the performance of all developed models is evaluated using the Apartment 2 test dataset. Table~\ref{tab:Performance_Apt2} summarizes the results for three scenarios: using all available features ($\Tindoor$, $\RH$, $\COtwo$, $\COtwo$ slope for all models, with season information included only for LR and SVM), excluding $\Tindoor$ and $\RH$, and excluding $\COtwo$ and $\COtwo$ slope. Fig.~\ref{fig:confusion_matrix} presents the corresponding confusion matrices for all models.
As noted earlier, these evaluations are conducted to determine whether reasonable model performance can still be achieved when only a subset of sensors is available in practical applications. When the full feature set is used, all three models demonstrate strong and comparable performance across all evaluation metrics. Considering the F1 score, which balances both precision and recall, the LSTM model shows a slightly higher performance than LR and SVM. Moreover, LR and SVM still achieve competitive accuracy and AUC-ROC values, indicating reliable estimation capability even without the temporal modeling inherent to LSTM. When either $\RH$ and $\Tindoor$ or $\COtwo$ and its slope are removed, the LSTM approach again achieves the highest F1 score, although the performance gap remains small. Importantly, all models maintain acceptable performance when fewer features are available, supporting their applicability in practical settings where sensor availability may be limited. Nonetheless, the noticeable performance degradation observed when excluding $\COtwo$ and the $\COtwo$ slope indicates that these variables are the most significant features for occupancy estimation across all model types.
\begin{table*}[h!]
\centering
\begin{tabular}{lcccccc}
\hline
\textbf{Features} & \textbf{Models} & \textbf{Precision} & \textbf{Recall} 
& \textbf{F1} & \textbf{Accuracy} & \textbf{AUC ROC} \\
\hline

All features
 & \begin{tabular}{@{}c@{}}LR\\SVM\\LSTM\end{tabular}
 & \begin{tabular}{@{}c@{}}0.8396\\0.8099\\\textbf{0.8436}\end{tabular}
 & \begin{tabular}{@{}c@{}}0.8807\\\textbf{0.9133}\\0.8885\end{tabular}
 & \begin{tabular}{@{}c@{}}0.8597\\0.8585\\\textbf{0.8655}\end{tabular}
 & \begin{tabular}{@{}c@{}}\textbf{0.8412} \\0.8184 \\0.8333\end{tabular}
 & \begin{tabular}{@{}c@{}}\textbf{0.9164}\\0.9028\\0.9100\end{tabular} \\
\hline
Excluding $\RH$ and $\Tindoor$
 & \begin{tabular}{@{}c@{}}LR\\SVM\\LSTM\end{tabular}
 & \begin{tabular}{@{}c@{}}0.8602\\0.8189\\\textbf{0.9105}\end{tabular}
 & \begin{tabular}{@{}c@{}}0.8137\\\textbf{0.8635}\\0.8234\end{tabular}
 & \begin{tabular}{@{}c@{}}0.8363\\0.8406\\\textbf{0.8648}\end{tabular}
 & \begin{tabular}{@{}c@{}}0.8241\\0.8024\\\textbf{0.8446}\end{tabular}
 & \begin{tabular}{@{}c@{}}0.8964 \\0.8774\\\textbf{0.9108}\end{tabular} \\
\hline
Excluding $\COtwo$ and $\COtwo$ slope
 & \begin{tabular}{@{}c@{}}LR\\SVM\\LSTM\end{tabular}
 & \begin{tabular}{@{}c@{}}\textbf{0.7873}\\0.6032\\0.7327\end{tabular}
 & \begin{tabular}{@{}c@{}}0.6469\\\textbf{1.0000}\\0.8224\end{tabular}
 & \begin{tabular}{@{}c@{}}0.7103\\0.7525\\\textbf{0.7749}\end{tabular}
 & \begin{tabular}{@{}c@{}}0.7086\\0.6032\\ \textbf{0.7118}\end{tabular}
 & \begin{tabular}{@{}c@{}}0.6987\\\textbf{0.7283}\\0.7181\end{tabular} \\

\hline
\end{tabular}
\caption{Performance metrics for the LR, SVM, and LSTM models evaluated on Apartment~2 test dataset. The highest score within each metric category is highlighted in bold.}
\label{tab:Performance_Apt2}
\end{table*}
\subsection{Cross-apartment generalization}\label{sec:generalization}

\begin{figure}
    \centering
    \includegraphics[width=0.6\linewidth]{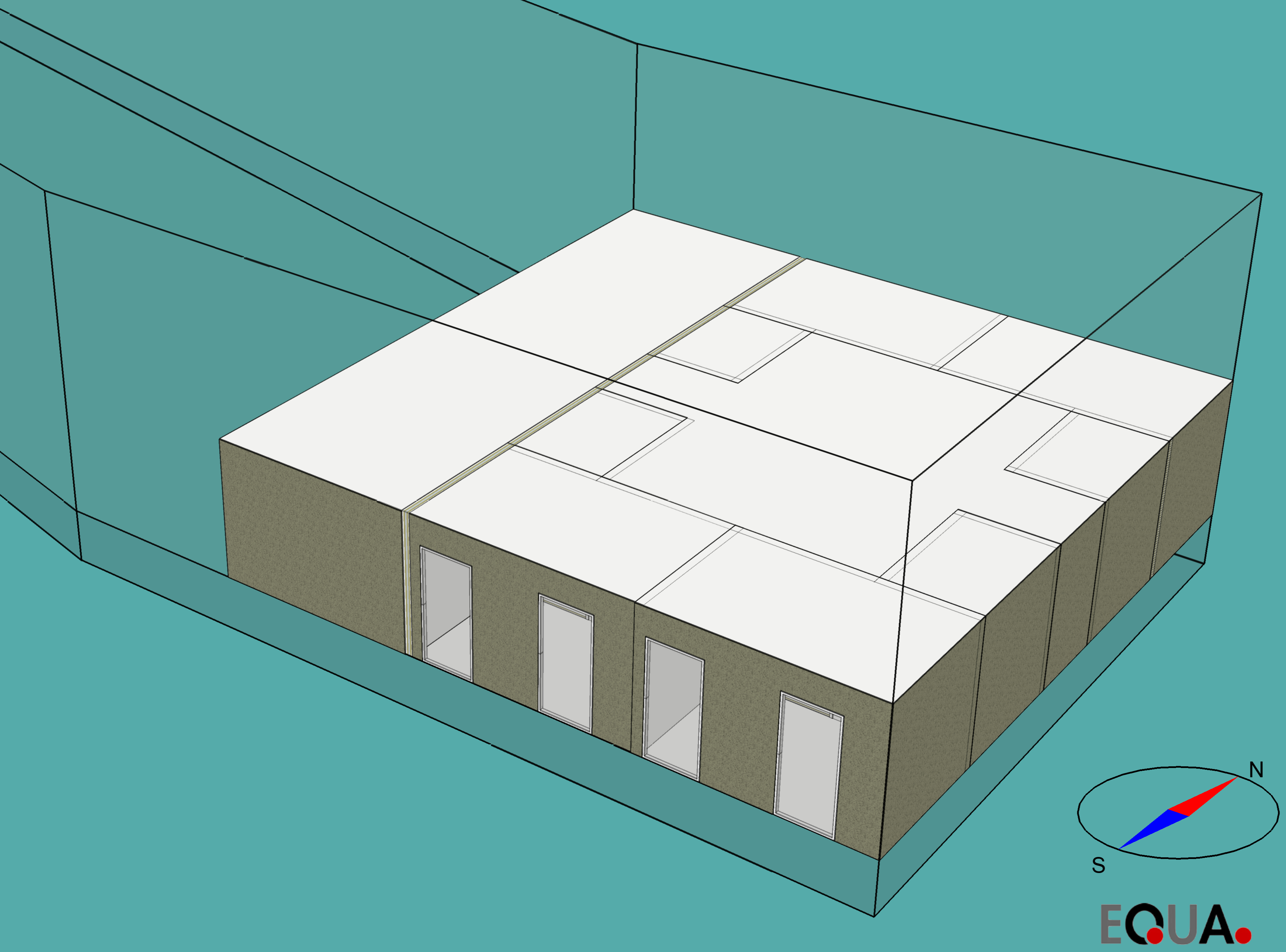}
    \caption{Testbed KTH model developed in IDA ICE. }
    \label{fig:TestbedKTH_IDAICE}
\end{figure}

\begin{figure}
    \centering
    \includegraphics[width=0.85\linewidth]{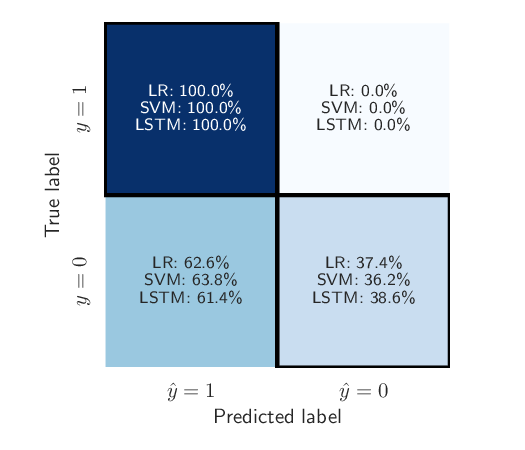}
    \caption{Row-normalized confusion matrices for the models evaluated on the digital model dataset.}
    \label{fig:confusion_matrix_IDAICE}
\end{figure}

To assess the generalizability of the developed models, two evaluation scenarios are considered.
\vspace{-0.4em}
\paragraph*{Scenario 1.}
\vspace{-1em}
The test data are generated using a calibrated digital model of Testbed KTH (Fig.~\ref{fig:TestbedKTH_IDAICE}), developed in the IDA ICE environment (see \citep{farjadnia2025towards} for details), an advanced, dynamic, multi-zone building simulation tool capable of producing high-fidelity representations of building performance \citep{kalamees2004ida}. In this setup, occupant-related factors (e.g., number of occupants, window and door operations) are fully controlled, enabling the generation of data under well-defined and reproducible conditions. Table \ref{tab:performance_IDAICEdata} and Fig.~\ref{fig:confusion_matrix_IDAICE} summarize model performance and the corresponding confusion matrices for the digital model dataset. The results indicate high performance across all models, with LR performing slightly better. This is likely because the synthetic dataset is less noisy than the real apartment data, allowing a low-complexity linear model such as LR to capture the dominant occupancy patterns effectively.

\begin{figure}
    \centering
    \includegraphics[width=0.85\linewidth]{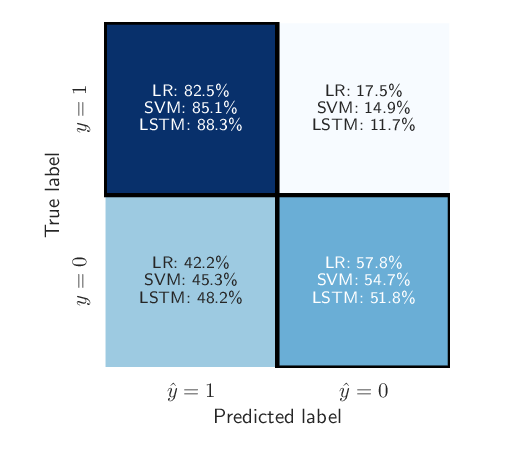}
    \caption{Row-normalized confusion matrices for the models evaluated on Apartment 3 data.}
    \label{fig:confusion_matrix_LGH3}
\end{figure}

\vspace{-1em}
\paragraph*{Scenario 2.} 
\vspace{-0.6em}
Three months of real experimental data from Apartment~3 of Testbed KTH are used. Unlike the digital model, this environment is not controlled and inherently includes occupant-driven uncertainties, providing a more realistic and challenging test for evaluating the robustness of the models under practical operating conditions. Table \ref{tab:performance_Apt3} and Fig.~\ref{fig:confusion_matrix_LGH3} present model performance and the corresponding confusion matrix for the Apartment~3 dataset. The results show reduced performance compared with the reference apartment and digital model datasets. However, all models achieve an acceptable F1 score, with the LSTM model performing best overall. 

\begin{table}[h!]
\centering
\begin{tabular}{lcccc}
\hline
\textbf{Model} & \textbf{Precision} & \textbf{Recall} & \textbf{F1} & \textbf{Accuracy} \\
\hline

LR   
 & \textbf{0.7235}
 & \textbf{1.0000}
 & \textbf{0.8396}
 &  \textbf{0.7626}\\

SVM  
 & 0.7198
 & 1.0000
 & 0.8371
 & 0.7582 \\

LSTM 
 & 0.7205
 & 1.0000
 & 0.8375
 & 0.7622 \\
\hline
\end{tabular}
\caption{Performance metrics for LR, SVM, and LSTM models trained on Apartment~2 data and evaluated on the digital model data.}
\label{tab:performance_IDAICEdata}
\end{table}

\begin{table}[h!]
\centering
\begin{tabular}{lcccc}
\hline
\textbf{Model} & \textbf{Precision} & \textbf{Recall} & \textbf{F1} & \textbf{Accuracy} \\
\hline

LR   
 & 0.6817
 & 0.8250
 & 0.7466
 & 0.7072 \\

SVM  
 & 0.6732
 & 0.8512
 & 0.7518
 & 0.7062 \\

LSTM 
 & \textbf{0.8436}
 & \textbf{0.8885}
 & \textbf{0.8655}
 &\textbf{0.8333} \\
\hline
\end{tabular}
\caption{Performance metrics for LR, SVM, and LSTM models trained on Apartment~2 dataset and evaluated on Apartment~3 data.}
\label{tab:performance_Apt3}
\end{table}

To compare the computational complexity of the models, the training time of each classifier is reported in Table~\ref{tab:runtime_models}.
\begin{table}[h!]
\centering
\begin{tabular}{lccc}
\hline
\textbf{Model} & \textbf{Time [s]} & \textbf{Time [min]} & \textbf{Relative to LR} \\
\hline

LR   
 & 11.3 
 & 0.19 
 & $1\times$ \\

SVM  
 & 2058.1
 & 34.3
 & $\approx 182\times$ \\

LSTM 
 & $>3000$
 & $>50$
 & $>265\times$ \\
\hline
\end{tabular}
\caption{Empirical training time and relative computational cost of LR, SVM, and LSTM models, considering LR as the $1\times$ baseline.}
\label{tab:runtime_models}
\end{table}
As listed in Table~\ref{tab:runtime_models}, LR is the least computationally expensive model, with a training time of $11.3$\,s, and is therefore used as a baseline with a relative cost of $1\times$. The LSTM model is the most expensive model, with a training time exceeding $50$\,min. Although it is the most computationally demanding classifier in this study, its cost is justified by its superior generalization to unseen data, as shown previously in Table~\ref{tab:performance_Apt3}. In this case, the additional training time translates into improved predictive performance rather than overfitting. However, if the primary objective is to estimate occupancy within the same building, the LR model offers acceptable performance while providing a substantially lower computational cost.

\section{Conclusion}\label{sec:Conclusion}
This paper evaluated three occupancy detection models, LR, SVM, and LSTM, using residential data from KTH Live-In Lab. The hyperparameters of both the SVM and LSTM methods were automatically optimized using Bayesian optimization. The results showed that all models achieved comparable performance when tested on data from the same apartment, while the LSTM model demonstrated the strongest generalization capability when evaluated on data from other apartments in the testbed, as well as on data from a calibrated digital model. It is important to note that LSTM and SVM have higher computational complexity than LR. Consequently, LR represents a viable low-complexity option for applications confined to single apartment occupancy detection, whereas LSTM is better suited for deployments requiring cross-apartment generalization. Future work will extend these models to jointly estimate occupancy and window-operation behavior \citep{farjadnia2023influences} and evaluate their combined impact on building energy consumption using the digital Testbed KTH model \citep{farjadnia2026assessing}.

\bibliography{ifacconf}             

@article{drgovna2020all,
  title={All you need to know about model predictive control for buildings},
  author={Drgo{\v{n}}a, J{\'a}n and Arroyo, Javier and Figueroa, Iago Cupeiro and Blum, David and Arendt, Krzysztof and Kim, Donghun and Oll{\'e}, Enric Perarnau and Oravec, Juraj and Wetter, Michael and Vrabie, Draguna L and others},
  journal={Annual Reviews in Control},
  volume={50},
  pages={190--232},
  year={2020},
  publisher={Elsevier}
}

@misc{IEA_Building,
  author       = {{International Energy Agency}},
  title        = {Buildings: Energy System},
  year         = {2022},
  note         = {\url{https://www.iea.org/energy-system/buildings}},
}

@article{molinari2023using,
  title={{Using living labs to tackle innovation bottlenecks: the KTH Live-In Lab case study}},
  author={Molinari, Marco and Vogel, Jonas Anund and Rolando, Davide and Lundqvist, Per},
  journal={Applied Energy},
  volume={338},
  number={120877},
  year={2023},
  publisher={Elsevier}
}

@article{rolando2022long,
  title={Long-term evaluation of comfort, indoor air quality and energy performance in buildings: The case of the {KTH Live-In Lab Testbeds}},
  author={Rolando, Davide and Mazzotti Pallard, Willem and Molinari, Marco},
  journal={Energies},
  volume={15},
  number={14},
  pages={4955},
  year={2022},
  publisher={MDPI}
}

@article{cali2016analysis,
  title={{Analysis of occupants' behavior related to the use of windows in German households}},
  author={Cal{\`\i}, Davide and Andersen, Rune Korsholm and M{\"u}ller, Dirk and Olesen, Bjarne W},
  journal={Building and Environment},
  volume={103},
  pages={54--69},
  year={2016},
  publisher={Elsevier}
}

@article{xu2023critical,
  title={A critical review of occupant energy consumption behavior in buildings: How we got here, where we are, and where we are headed},
  author={Xu, Xiaoxiao and Yu, Hao and Sun, Qiuwen and Tam, Vivian WY},
  journal={Renewable and Sustainable Energy Reviews},
  volume={182},
  pages={113396},
  year={2023},
  publisher={Elsevier}
}

@article{rueda2020comprehensive,
  title={A comprehensive review of approaches to building occupancy detection},
  author={Rueda, Luis and Agbossou, Kodjo and Cardenas, Alben and Henao, Nilson and Kelouwani, Sousso},
  journal={Building and Environment},
  volume={180},
  pages={106966},
  year={2020},
  publisher={Elsevier}
}

@article{amiri2025review,
  title={A review of occupancy detection techniques for {HVAC} control: Advances and practical challenges},
  author={Amiri, Alireza Jafarian and Mahmoodi, Kaywan and Tootchi, Amirreza and Razban, Ali},
  journal={Journal of Building Engineering},
  pages={113962},
  year={2025},
  publisher={Elsevier}
}

@article{park2019critical,
  title={A critical review of field implementations of occupant-centric building controls},
  author={Park, June Young and Ouf, Mohamed M and Gunay, Burak and Peng, Yuzhen and O'Brien, William and Kj{\ae}rgaard, Mikkel Baun and Nagy, Zoltan},
  journal={Building and Environment},
  volume={165},
  pages={106351},
  year={2019},
  publisher={Elsevier}
}

@article{jin2021building,
  title={Building occupancy forecasting: A systematical and critical review},
  author={Jin, Yuan and Yan, Da and Chong, Adrian and Dong, Bing and An, Jingjing},
  journal={Energy and Buildings},
  volume={251},
  pages={111345},
  year={2021},
  publisher={Elsevier}
}

@article{shi2017energy,
  title={Energy efficient building {HVAC} control algorithm with real-time occupancy prediction},
  author={Shi, Jie and Yu, Nanpeng and Yao, Weixin},
  journal={Energy Procedia},
  volume={111},
  pages={267--276},
  year={2017},
  publisher={Elsevier}
}

@article{chen2017environmental,
  title={Environmental sensors-based occupancy estimation in buildings via {IHMM-MLR}},
  author={Chen, Zhenghua and Zhu, Qingchang and Masood, Mustafa Khalid and Soh, Yeng Chai},
  journal={IEEE Transactions on Industrial Informatics},
  volume={13},
  number={5},
  pages={2184--2193},
  year={2017},
  publisher={IEEE}
}

@article{wang2018occupancy,
  title={Occupancy prediction through machine learning and data fusion of environmental sensing and {Wi-Fi} sensing in buildings},
  author={Wang, Wei and Chen, Jiayu and Hong, Tianzhen},
  journal={Automation in Construction},
  volume={94},
  pages={233--243},
  year={2018},
  publisher={Elsevier}
}

@article{kim2019real,
  title={Real-time occupancy prediction in a large exhibition hall using deep learning approach},
  author={Kim, Seonghyeon and Kang, Seokwoo and Ryu, Kwang Ryel and Song, Giltae},
  journal={Energy and Buildings},
  volume={199},
  pages={216--222},
  year={2019},
  publisher={Elsevier}
}

@article{dong2010information,
  title={An information technology enabled sustainability test-bed {(ITEST)} for occupancy detection through an environmental sensing network},
  author={Dong, Bing and Andrews, Burton and Lam, Khee Poh and H{\"o}ynck, Michael and Zhang, Rui and Chiou, Yun-Shang and Benitez, Diego},
  journal={Energy and Buildings},
  volume={42},
  number={7},
  pages={1038--1046},
  year={2010},
  publisher={Elsevier}
}

@inproceedings{khalil2021transfer,
  title={Transfer learning approach for occupancy prediction in smart buildings},
  author={Khalil, Mohamad and McGough, Stephen and Pourmirza, Zoya and Pazhoohesh, Mehdi and Walker, Sara},
  booktitle={12th International renewable engineering conference (IREC)},
  pages={1--6},
  year={2021},
  organization={IEEE}
}

@article{candanedo2017methodology,
  title={A methodology based on Hidden Markov Models for occupancy detection and a case study in a low energy residential building},
  author={Candanedo, Luis M and Feldheim, V{\'e}ronique and Deramaix, Dominique},
  journal={Energy and Buildings},
  volume={148},
  pages={327--341},
  year={2017},
  publisher={Elsevier}
}

@article{liang2024low,
  title={Low-cost data-driven estimation of indoor occupancy based on carbon dioxide {(CO2)} concentration: A multi-scenario case study},
  author={Liang, Xiguan and Shim, Jisoo and Anderton, Owen and Song, Doosam},
  journal={Journal of Building Engineering},
  volume={82},
  pages={108180},
  year={2024},
  publisher={Elsevier}
}

@article{azimi2022fit,
  title={Fit-for-purpose: Measuring occupancy to support commercial building operations: A review},
  author={Azimi, Sara and O'Brien, William},
  journal={Building and Environment},
  volume={212},
  pages={108767},
  year={2022},
  publisher={Elsevier}
}

@article{andersen2024exploring,
  title={Exploring occupant detection model generalizability for residential buildings using supervised learning with IEQ sensors},
  author={Andersen, Kamilla Heimar and Johra, Hicham and Schaffer, Markus and Marszal-Pomianowska, Anna and Knudsen, Henrik N and Heiselberg, Per Kvols and O'Brien, William},
  journal={Building and Environment},
  volume={254},
  pages={111319},
  year={2024},
  publisher={Elsevier}
}

@article{kalamees2004ida,
  title={{IDA ICE}: the simulation tool for making the whole building energy and HAM analysis},
  author={Kalamees, Targo},
  journal={Annex},
  volume={41},
  pages={12--14},
  year={2004}
}

@phdthesis{farjadnia2025towards,
  title        = {Towards Human-in-the-Loop Smart Buildings: Data-Driven Predictive Control and Occupant Modeling},
  author       = {Farjadnia, Mahsa},
  year         = {2025},
  type         = {Licentiate Thesis},
  school       = {Kungliga Tekniska h{\"o}gskolan},
  address      = {Stockholm, Sweden}
}

@book{hosmer2013applied,
  title={Applied logistic regression},
  author={Hosmer Jr, David W and Lemeshow, Stanley and Sturdivant, Rodney X},
  year={2013},
  publisher={John Wiley \& Sons}
}

@inproceedings{farjadnia2023influences,
  title={What influences occupants’ behavior in residential buildings: An experimental study on window operation in the {KTH Live-In Lab}},
  author={Farjadnia, Mahsa and Fontan, Angela and Russo, Alessio and Johansson, Karl Henrik and Molinari, Marco},
  booktitle={2023 IEEE Conference on Control Technology and Applications (CCTA)},
  pages={752--758},
  year={2023},
  organization={}
}

@INPROCEEDINGS{Snoek2012,
 author = {Snoek, Jasper and Larochelle, Hugo and Adams, Ryan P},
 booktitle = {Advances in Neural Information Processing Systems},
 title = {Practical Bayesian Optimization of Machine Learning Algorithms},
 volume = {25},
 year = {2012}
}

@misc{ba2016layernormalization,
      title={Layer Normalization}, 
      author={Jimmy Lei Ba and Jamie Ryan Kiros and Geoffrey E. Hinton},
      year={2016},
      eprint={1607.06450},
      archivePrefix={arXiv}
}

@article{farjadnia2026assessing,
  title={Assessing the impact of occupant behavior on residential building performance: A case study of window operation},
  author={Farjadnia, Mahsa and Fontan, Angela and Johansson, Karl Henrik and Molinari, Marco},
  journal={Building and Environment},
  pages={114552},
  year={2026},
  publisher={Elsevier}
}

@article{li2017new,
  title={A new modeling approach for short-term prediction of occupancy in residential buildings},
  author={Li, Zhaoxuan and Dong, Bing},
  journal={Building and Environment},
  volume={121},
  pages={277--290},
  year={2017},
  publisher={Elsevier}
}

@article{huchuk2019comparison,
  title={Comparison of machine learning models for occupancy prediction in residential buildings using connected thermostat data},
  author={Huchuk, Brent and Sanner, Scott and O'Brien, William},
  journal={Building and Environment},
  volume={160},
  pages={106177},
  year={2019},
  publisher={Elsevier}
}
\end{document}